\documentstyle[12pt,epsf]{article}
\begin{document}
\begin{titlepage}
  \begin{flushright}
    KUNS-1698\\[-1mm]
    hep-ph/0011374
  \end{flushright}
  \begin{center}
    \vspace*{1.4cm}
    
    {\Large\bf Brane Fluctuation and New Counting Rules\\ for
    Kaluza-Klein Towers} 
    \vspace*{1cm}
  
    Masako Bando\footnote{E-mail: bando@aichi-u.ac.jp} and
    Tatsuya Noguchi\footnote{E-mail: noguchi@gauge.scphys.kyoto-u.ac.jp}
    \vspace*{5mm}
  
  $^*${\it Aichi University, Aichi 470-0296, Japan}\\
  $^\dagger${\it Department of Physics, Kyoto
    University, Kyoto 606-8502, Japan}
  \vspace{1.2cm}
  
  \begin{abstract} In models with extra dimensions, branes have been
   usually treated as solid bodies though they are prohibited by the
   relativity.  In the previous letter, we proposed a method of taking
   account of brane fluctuation by introducing Nambu-Goldstone bosons,
   and prove that when a tension of the brane is small, the
   interaction between boundary fields and Kaluza-Klein modes is
   suppressed exponentially.  In this letter, we further investigate
   this suppression in more generic configuration, and obtain three
   counting rules, ``AND'', ``OR'' and ``STACK'' rules, depending on
   the softness of branes and the character of fields on the branes
   four dimensions. The choice determines the number of Kaluza-Klein
   towers contributing to renormalization group equations, leading to
   a remarkable change in the running behavior of coupling constants.
  \end{abstract}
\end{center}
\end{titlepage}

\setcounter{footnote}{0}

\section{Introduction}
\setcounter{equation}{0}

Over the past years models with extra dimensions have been intensively
studied from various viewpoints \cite{base}.  Within the framework of
these models, our spacetime is thought to be a four dimensional object
located in higher dimensional space, and observable quarks and leptons
are regarded as either fermion fields defined on the brane or
Kaluza-Klein zero modes of bulk fermion fields.  Gauge and
gravitational fields are usually assumed to propagate in the bulk
space.  Therefore, the interaction between boundary and bulk fields
plays an important r\^ole when it comes to discussing phenomenology.
When constructing Lagrangians with boundary-bulk interactions, we have
usually treated the boundary as a membranous solid body.  However, the
fact that the relativity prohibits the existence of solid bodies
suggest that we should take account of the effect of brane
fluctuation.

In the previous letter, we propose a method of incorporating brane
fluctuation by introducing Nambu-Goldstone bosons originated from the
spontaneous breaking of the translational symmetry along extra spatial
directions \cite{brane-fluctuation}.  Consequently, we show the
exponential suppression of the interaction between four dimensional
electron fields and Kaluza-Klein modes of five dimensional photon
fields when the brane tension is sufficiently small.\footnote{Similar
suppression factor was obtained in a different context in
Ref.~\cite{sup}.}  The analysis can be immediately extended to the
interaction involving generic boundary-boundary-bulk vertices.  If we
adopt this prescription, we do not have to introduce an ultraviolet
cutoff required to regulate the divergence from an infinite summation
of Kaluza-Klein modes because interactions with higher Kaluza-Klein
modes are properly suppressed. While the smallness of the interaction
unfortunately lessens the possibility of observing extra dimensions,
on the other hand it provides an interesting phenomenon of observing
the scalar fields describing brane fluctuation \cite{kugo-yoshioka}.

In this letter, we will further investigate the effect of brane
fluctuation in more generic situations where fields involving
boundary-bulk interactions extend in different dimensional spaces, and
according to the softness of the branes we obtain three counting rules
for Kaluza-Klein towers, which are provisionally called as ``AND'',
``OR'' and ``STACK'' rules.

More interestingly, the alteration of counting rules drastically
changes a running behavior of coupling constants because it is the
number of Kaluza-Klein towers propagating in a relevant loop that
determines an exponent of the power-law
\cite{ddg}\cite{bkny1}\cite{bkny2}. 

\section{Solid brane system}
\setcounter{equation}{0}

In this section, we briefly review usual treatment for boundary-bulk
interaction under the solid body approximation, and show the resultant
power-law behavior of beta functions \cite{ddg}.

Within models with extra dimensions, matter fields such as quarks and
leptons are regarded as (i) fermion fields defined only on a brane, or
(ii) Kaluza-Klein zero modes of bulk fermion fields. For simplicity,
we assume the number of the extra spatial dimension to be one, the
generalization being straightforwardly to the case with more extra
spatial dimensions.  In order for our discussion to be strict, we
consider only scalar fields as boundary and bulk fields throughout
this article. We expect that one may perform the same analysis with
fermion fields and generalize our setup to supersymmetric models.

Let us investigate a concrete interaction involving with a scalar
field $\phi_1(x)$ satisfying the case (i) and $\phi_2(x,y)$ and
$\phi_3(x,y)$ propagating in five dimensional spacetime. In this
model, an interaction term can be described as
\begin{eqnarray}
\label{interacting_action}
S_{\rm int} = \int d^4x\, \int_0^{2\pi R} dy\, Y
\phi_1(x)\,\delta(y)\,\phi_2(x,y)\phi_3(x,y).
\end{eqnarray}
where $Y$ is a coupling constant and $\delta(y)$ is inserted in order
to take the boundary values of the bulk fields.  Decomposing $\phi_2$
and $\phi_3$ into Kaluza-Klein modes $\phi_2 = \sum_n
\phi_2^{(n)}(x)e^{iny/R}$ and $\phi_3 = \sum_m
\phi_3^{(m)}(x)e^{imy/R}$, the equation (\ref{interacting_action}) is
rewritten in terms of four dimensional field theory as
\begin{eqnarray}
S_{\rm int} = \int d^4x\, Y \sum_{n,m=-\infty}^\infty \phi_1(x)\,
\phi_2^{(n)}(x)\, \phi_3^{(m)}(x).
\label{solidbrane}
\end{eqnarray}

Since all Kaluza-Klein modes interact with $\phi_1(x)$ with an equal
coupling constant, the Kaluza-Klein modes of $\phi_2(x,y)$ and
$\phi_3(x,y)$ contribute equally to renormalization group equations.
We should note that this result comes from the $\delta(y)$ function in
Eq.~(\ref{interacting_action}).  In addition, owing to the
localization of $\phi_1(x)$ in the higher dimensional space, we cannot
impose a momentum conservation rule along extra spatial directions,
thus there appears no restriction about the Kaluza-Klein excitation
numbers $n$ and $m$. Therefore, two Kaluza-Klein towers parameterized
by integers $n$ and $m$ contribute independently to the anomalous
dimension of $\phi_1(x)$, and finally we obtain the power-law
dependence with the exponent 2.

On the other hand, if the boundary field $\phi_1(x)$ is assumed to be
a Kaluza-Klein zero mode, the action can be written as
\begin{eqnarray}
S_{\rm int} = \int d^4x\, \int_0^{2\pi R} dy\, Y
\phi_1^{(0)}(x)\,\phi_2(x,y)\phi_3(x,y).
\label{solid_action}
\end{eqnarray}
In terms of four dimensions, we obtain
\begin{eqnarray}
S_{\rm int} = \int d^4x\, Y\sum_{n,m=-\infty}^\infty \phi_1^{(0)}(x)
\phi_2^{(n)}(x) \phi_3^{(m)}(x)e^{\frac{i(n+m)y}{R}}.
\label{solidii}
\end{eqnarray}
The assumption that $\phi_1(x)$ is a Kaluza-Klein zero mode requires a
momentum conservation rule along the extra spatial direction, thus
only the Kaluza-Klein modes satisfying $n+m=0$ contribute to
renormalization group equations. Consequently, we obtain the power-law
behavior with the exponent 1.  It should be noted that the case (ii)
implies a momentum conservation rule along extra spatial directions
while the case (i) does not because their extra dimensional momenta
are indefinite.

Similarly, it is found that when both $\phi_1$ and $\phi_3$ are
boundary fields, the resultant exponent is always one whether they
belong to the case (i) or (ii).

Here we recapitulate these results:
\begin{itemize}
\item in the case with $\phi_1$ belonging to the type (i) fields, the
    exponent turns out to be the number of fields propagating to the
    extra direction in a relevant loop;
\item in the case with $\phi_1$ belonging to the type (ii) fields, the
  exponent turns out to be one if at least one of the fields $\phi_2$
  and $\phi_3$ expands in the extra dimension; otherwise zero.
\end{itemize}
We can put these rules in the following way.  As for the former,
{\it the total number of Kaluza-Klein towers propagating in a relevant
  loop} contributes to the exponent.
On the other hand, as for the latter, only
{\it the extra dimension where $\phi_2(x)$ or $\phi_3(x)$ extends}
contributes to the total exponent of the power-law. 
 Therefore, we call the former ``STACK'' rule and the latter
``OR'' rule.  In Ref.~\cite{bkny2}, we showed that the OR rule
prefers democratic Yukawa matrices.\footnote{In the paper, we did not
call this rule as ``OR'' rule explicitly.}  On the basis of STACK
rule, the exponent of beta functions of Yukawa couplings differs
generically depending on their generation, thus it seems viable to
realize fermion mass hierarchies \cite{bkny2}.  Further investigation
on this point has been in progress \cite{phenomenology}.

In the case with more extra dimensions, the total exponent would be
given by the sum of the contribution from each extra direction. 

\section{Soft brane system}
\setcounter{equation}{0}

Let us follow the prescription proposed in
Ref.~\cite{brane-fluctuation} and introduce a Nambu-Goldstone boson
$\phi(x)$ stemming from the spontaneous breaking of extra dimensional
translation symmetry. 

The Lagrangian of this system remains the same as the preceding one
(\ref{interacting_action}), but there appears a kinetic term of
$\phi(x)$ from a determinant of the induced metric, i.e.~the Nambu-Goto 
action $\int d^4x(-\tau_4)\sqrt{-g}$,
\begin{eqnarray}
  \int d^4 x\left\{-\tau_4 +
  \frac{\tau_4}{2}\partial_\mu\phi(x)\partial^\mu\phi(x) + \cdots\right\},\label{kin}
\end{eqnarray}
where $\tau_4$ is a four dimensional tension and the ellipsis stands
for negligible higher derivative terms.

By introducing this scalar fields $\phi(x)$ we here take account of
brane fluctuation; we substitute $y$ in Eq.~(\ref{solid_action}) with
$\phi(x)$ and obtain the following boundary-bulk interaction \cite{brane-fluctuation},
\begin{eqnarray}
S_{\rm int} &=& \int d^4x\,Y\sum_{n,m}\phi_1(x)\phi_2^{(n)}(x)
\phi_3^{(m)}(x)e^{\frac{i(n+m)\phi(x)}{R}}  
\end{eqnarray}
Since our analysis are based on a perturbation theory, it seems
appropriate to rewrite the exponential factor
$\exp(i\frac{n}{R}\phi(x))$ into the normal ordered form referring to
the free kinetic term of $\phi$ in Eq.~(\ref{kin}).
\begin{eqnarray}
S_{\rm int}&\sim& \int
d^4x\,Y\sum_{n,m}\phi_1(x)\phi_2^{(n)}(x)\phi_3^{(m)}(x)
e^{-\frac{1}{2}\left(\frac{n+m}{R}\right)^2\Delta(l_s)} :
e^{\frac{i(n+m)\phi(x)}{R}}:\label{normal-ordered}
\end{eqnarray}
where $\Delta$ is the free propagator of $\phi$ defined as
\begin{eqnarray}
  \Delta(x-y) \equiv \langle\phi(x)\phi(y)\rangle =
  \frac{1}{f^4}\cdot\frac{1}{-(x-y)^2},
\end{eqnarray}
Note that $f^{-1}$ is a characteristic length of brane fluctuation
along the extra spatial direction and related to the brane tension
through $\tau_4 = \frac{f^4}{4\pi^2}$.  The limit $f\rightarrow
\infty$ corresponds to the solid brane approximation.  Since we have
treated this system as an effective theory valid below the cutoff
energy scale $M_s = l_s^{-1}$, the propagator $\Delta(x)$ with $|x|
\le l_s$ should be regarded as $\Delta(l_s) = 1/(f^4l^2_s)$, and thus
we have replaced the infinite $\Delta(0)$ by the value $\Delta(l_s)$
in Eq.~(\ref{normal-ordered}).  We can read the effective Yukawa
coupling $Y_{n,m}$ from Eq.~(\ref{normal-ordered}),
\begin{eqnarray}
  Y_{n,m} =
  e^{-\frac{1}{2}\left(\frac{n+m}{R}\right)^2\frac{M_s^2}{f^4}}\,Y.
\end{eqnarray}

In the case with a small tension, this effective coupling signifies
apparently that only the diagrams satisfying the relation $n + m=0$
contribute to the anomalous dimension.  This restriction on the
combination of momenta can be effectively interpreted as an extra
dimensional momentum conservation rule.  If we take the limit of
$f\rightarrow\infty$, the suppression factor of the effective Yukawa
coupling disappears and we are led to the same coupling shown in the
solid body system.

This effective momentum conservation rule can be interpreted
intuitively as follows.  In the case with a small brane tension, when
a boundary field decays into a boundary field and only one
Kaluza-Klein mode of bulk fields, the brane gets deformed by the
back-reaction. Accordingly, the overlapping between the wave function
of the initial state and that of the final state (i.e.~before and
after the emission) gets suppressed exponentially. On the other hand,
when a boundary field decays into two Kaluza-Klein modes carrying
equal momenta along the extra spatial directions with an opposite
sign, the brane remains unchanged.  As a result, the coupling does not
suffer the exponential suppression.  This is the reason why only one
Kaluza-Klein tower contributes to the wave functional renormalization
of $\phi_1(x)$.

In order to complete the classification of our counting rules let us
finally investigate the case that the scalar field $\phi_3$ does not
extend in the extra spatial direction, i.e.~$\phi_3$ is defined only
on the four dimensional brane (case (i)) or regarded as a Kaluza-Klein
zero mode (case (ii)).  In this configuration, whether $\phi_3$
belongs to the case (i) or (ii), the action is written as
\begin{eqnarray}
S_{\rm int} &=& \int
d^4x\,Y\sum_{n,m}\phi_1(x)\phi_2^{(n)}(x)\phi_3(x)e^{-{\left(\frac{n}{R}\right)}^2
\frac{M_s^2}{f^4}}
\end{eqnarray}
 and it indicates that the couplings between Kaluza-Klein modes of
$\phi_3$ and $\phi_1$ is suppressed for the case with a small brane
tension.

Taking account of the results obtained in the case with a soft brane,
 we find that it is only {\it an extra spatial dimension where both
 $\phi_2$ and $\phi_3$ extend} that contributes to an anomalous
 dimension of $\phi_1$, thus let us call this counting rule as ``AND''
 rule. \footnote{``OR'' and ``AND''$\,$ in the name of the counting
 rules are analogous to those in the logic circuit and the boolean
 algebra. We would observe the analogy explicitly in the table
 \ref{table1} after regarding ``bulk'' and ``boundary'' as ``1'' and
 ``0'', respectively.}  In table \ref{table1}, we summarize the
 counting rules for various cases.

\begin{table}[ht]
\begin{center}
\begin{tabular}{|c|c|c|c|c|}
\hline $\phi_1$ & \quad $\;\;\phi_2^{(m)}\;\;$ \qquad &
$\phi_3^{(n){\rm\,or\,}\o}$ & \# of KK towers\\ \hline\hline solid
brane & bulk & bulk & $2$ \\ \hline solid brane & bulk & boundary &
$1$ \\ \hline\hline KK zero mode & bulk & bulk & $1$ \\ \hline KK zero
mode & bulk & boundary & $1$ \\ \hline\hline fluc.~brane & bulk & bulk
& $1$ \\ \hline fluc.~brane & bulk & boundary & $0$ \\ \hline
\end{tabular}
\caption{The number of Kaluza-Klein towers contributing to the wave
 function renormalization of $\phi_1$ and the configuration of
 relevant fields. The number is determined by the property of the
 fields $\phi_1$, an extra spatial configuration of $\phi_2$ and
 $\phi_3$ and a brane tension. Note that the index $\o$ in the first
 line of the table means that the fields $\phi_3$ does not extend in
 the extra spatial dimension.}
\label{table1}
\end{center}
\end{table}
\begin{table}[ht]
\begin{center}
\begin{tabular}{|c|c|c|c|c|c|}
\hline \phantom{\qquad} $\phi_1$ \phantom{\qquad} &
\phantom{\qquad}counting rules\phantom{\qquad}\\ \hline\hline solid
brane & STACK \\ \hline KK zero mode & OR \\ \hline fluc.~brane & AND
\\ \hline
\end{tabular}
\end{center}
\caption{The relation between the property of $\phi_1$ and the
  resultant counting rule.}
\label{table2}
\end{table}

\section{Conclusion}
\setcounter{equation}{0}

We have investigated the effect of brane fluctuation on boundary-bulk
interactions on the basis of the method proposed in the previous
paper, and obtained three counting rules for Kaluza-Klein towers
contributing to renormalization group equations. We have shown that
one should adopt an appropriate counting rule judging from a brane
tension and configuration of scalar fields when it comes to evaluating
Feynman diagrams.  It should be noted that the alteration of counting
rules changes the exponents of renormalization group equations, and
thus occurs a drastic change in the running behavior of coupling
constants, which plays a significant r\^ole in discussing
phenomenology. Further investigation on this point will be shown in
our future paper \cite{phenomenology}.

Finally, we here make a brief comment on realization of chiral matter
fields. As is well known, the standard model fermions such as quarks
and leptons have been correctly treated as chiral multiplets. Within
the framework of field theories with extra dimensions it is known that
such a simple compactification as on a torus unfortunately cannot
produce chiral fermions, and that one must take more complicated
procedure such as orbifold compactifications with an appropriate
projection, where desirable chiral fermions appear on the fixed brane
in the whole spacetime. For example, in the effective description of
heterotic M-theory they appear on the fixed brane of the
$S^1/Z_2\times M^4$ manifold after $Z_2$ projection
\cite{horava-witten}\cite{projection}.  In general, such an orbifold
projection with respect to extra dimensions suggests that the fixed
brane should be treated as a rigid object, and the incorporation of
the brane fluctuation seems naively incompatible with a chiral
projection.

However, since it is clear that no rigid brane exists in the
relativistic theory, we expect that even the orbifold brane should
fluctuate so far as the tension is finite, where chiral fermions can
be realized on the fluctuating brane after an appropriate projection.
Thus we expect that the similar counting rules can be applied also to
the system with chiral fermions.  Further investigations should be
performed in order to confirm the validity, which is our next task to
be accomplished.

\section*{Acknowledgments}

The authors would like to thank T.~Kugo, H.~Nakano, T.~Kobayashi and
K.~Yoshioka for valuable discussions and comments. M.~B.~is supported
in part by the Grants-in-Aid for Scientific Research No.~12047225(A2)
and 12640295(C2) from the Ministry of Education, Science, Sports and
Culture, Japan.

\end{document}